\documentclass[letterpaper, a4paper]{amsart}

\usepackage{amsmath}
\usepackage{latexsym}
\usepackage[all]{xy}
\usepackage{color}

\include{amslatex}

 \numberwithin{equation}{section}

\begin{document}
\begin{title}[Maximal acceleration and black hole evaporation]
{Maximal acceleration and black hole evaporation}
\end{title}
\maketitle
\begin{center}
\author{Ricardo Gallego Torrom\'e\footnote{email: rigato39@gmail.com}}
\end{center}

\begin{center}
\address{Department of Mathematics\\
Faculty of Mathematics, Natural Sciences and Information Technologies\\
University of Primorska, Koper, Slovenia}
\end{center}

\begin{abstract}
We discuss how in certain theories of spacetime admitting a maximal proper acceleration Hawking radiation does not completely evaporate the black hole. The black hole remnant's mass depends on the inverse of the maximal acceleration. Furthermore, as a consequence of a duality between the minimum mass that a Schwarzschild black hole can have in such spacetimes and the maximal acceleration, we show that in certain theories with a maximal acceleration there must be an upper uniform bounds on the value of the force that can be exerted on test particles.
\end{abstract}
\section{Introduction}
The theoretical discovery of the black hole radiation and its consequences for the evaporation of black hole \cite{Hawking 1974} has posed a fundamental challenge, since the complete evaporation of the black hole seems to violate the unitary property of the quantum evolution dynamics. This problem has attracted continuous attention and multiple solutions have been proposed in the literature, as for instance it is discussed in   \cite{Giddings 1992,Giddings 1995,Nicolic}.

In this paper we show that certain classical geometric frameworks of spacetimes with a maximal proper acceleration imply a remnant to the black hole radiation process, supporting the possibility of a resolution of the information paradox in terms of the possibility that the remnant retains the {\it mixing information}. In general relativity, despite the related conjecture on the existence of a maximum force \cite{Gibbons 2002,Barrow Gibbons,Schiller 2005}, there is no limitation on the maximal proper acceleration that point test particles can experience. Therefore, it is necessary to go beyond general relativity to find frameworks with an upper uniform bound for the proper acceleration of test particles and observers. It is known that, in the ambit of quantum theories of gravity, there are arguments in favour of a maximal acceleration \cite{Bowick Giddins,Frolov Sanchez 1991,Parentani Potting,Rovelli Vidotto}.

In this paper we have shown the weak equivalence principle can be implemented in certain {\it high order jet spacetimes} compatible with the existence of a maximal acceleration. High order jet spacetimes geometries appear in the context of Caianiello's theory of maximal acceleration \cite{Caianiello,CaianielloFeoliGasperiniScarpetta} and in the context of violation of clock hypothesis in classical relativistic theories \cite{Gallego-Torrome2015}. These and other theories can be embraced in the formalism of {\it high order jet spacetimes of maximal acceleration} as envisaged in \cite{Gallego-Torrome2015}. Furthermore, the high order jet spacetimes that we will consider are not in principle related to quantum theories of gravity. In this context, we will argue that Fulling-Davies-Unruh theory can be applied, implying the existence of a maximal Unruh temperature from the existence of a maximal acceleration. This is a generalization of Cainiello and Landi's work in the context of quantum geometry \cite{CaianielloLandi}. It has also been shown how the weak equivalence principle can be implemented in high order jet spacetimes of maximal acceleration. With these three ingredients, namely, that the spacetime is furnished with  metrics of maximal acceleration, that the Unruh temperature formula holds good and the weak equivalence principle, one can show the existence of a maximal temperature for black holes of {\it Schwarzschild's type} and the existence of a minimal mass for such black holes.

The relation between an uniform upper bound in the temperature and its implication for the existence of a maximal upper bound in acceleration was first observed by H. Brandt in a particular framework and was used as a first motivation for the derivation of a maximal proper acceleration as a consequence of Sakharov's maximal temperature and Hawking temperature formula \cite{Brandt1983}. In this paper, however, we have considered an ample framework than in Brandt's original theory and investigated further the relation between maximal temperature and maximal proper acceleration, vastly generalizing Brandt's result and looking at it from the converse way.

One of the consequences of these considerations is the existence of a {\it duality relation} between the minimum mass that a Schwarzschild's type black hole can have in such geometries and the maximal acceleration that a test particle can experience. This relation leads to the existence of a maximum force exerted on test particles in the background of Schwarzschild's type black holes. The bound that we found for the force is the same as the one conjectured by Gibbons, Barrow and Schiller in the context of general relativity \cite{Gibbons 2002,Barrow Gibbons,Schiller 2005}, but the significance of our theory is rather different. Indeed, the argument shows that in the limit when the maximal acceleration tends to infinity, a limit case that holds for general relativity, the requirement of consistency of the duality relation mentioned above implies that there is no universal maximum force, suggesting the need to go beyond general relativity to find an uniform upper limit in the force.
\section{Maximal acceleration implies maximal temperature}
According to Unruh and others \cite{Unruh 1976, Wald 1994}, an accelerated observer feels itself immersed in a thermal bath whose temperature is given by the expression
\begin{align}
T=\,\frac{\hbar\, a}{2\pi\,c \, k_B},
\label{Unruh temperature}
\end{align}
 where $a$ is the modulus of the proper acceleration of a test particle or observer. As a consequence, if Unruh formula is still applicable in the setting of theories with a maximal proper acceleration, then the existence of an upper bound for maximal proper acceleration $A_{\textrm{max}}$ naturally implies a maximal temperature given by the expression
\begin{align}
T_{\textrm{max}}=\,\frac{\hbar\, A_{\textrm{max}}}{2\pi\,c \, k_B}.
\label{maximal temperature and acceleration}
\end{align}

We want to extend the validity of Unruh temperature formula to certain classical spacetimes with maximal acceleration. One can consider first, for instance, Brandt's and Cainiello's theories, which have a quantum mechanical motivation. The modification of the relativistic spacetime structure $(M,\eta)$ is a  {\it high order jet metric structures} that can be expressed in the form \cite{Brandt1989,CaianielloFeoliGasperiniScarpetta,Gallego-Torrome2007,Gallego-Torrome2015}
\begin{align}
g:=\Big(1+ \frac{  \eta(D_{x'}x'(\tau),
D_{x'}x'(\tau))}{{A}^2_{\textrm{max}}\,\eta(x',x')}\Big)\,\eta .
\label{maximalaccelerationmetric0}
\end{align}
Here the $'$-notation denotes the derivative with respect to the Lorentzian limit metric $g\to\,\eta$ as $A_{\textrm{max}}\to +\infty$. Such Lorentzian spacetime is denoted by $(M,\eta)$. Note that from the point of view of high order jet spacetime geometry \cite{Gallego-Torrome2015} and high order jet fields \cite{Ricardo2017b}, the spacetime structure is $g$, while $\eta$ can be seen as the limit structure corresponding to the limit of very small acceleration compared with $A_{\textrm{max}}$. $D_{x'}$ is the associated covariant derivative along the curve $x:I\to M,\,I\subset \mathbb{R}$ of the Levi-Civita connection of $\eta$. When using parameterizations such that $\eta(x',x')=\,-1$, the requirement of preserving the timelike character with respect to the metric $g$ and the metric $\eta$ implies that the proper acceleration $D_{\dot{x}}\dot{x}$ with respect to the metric $\eta$ is bounded by $A_{\textrm{max}}$,
\begin{align}
a^2=\,\eta(D_{x'}x'(\tau),
D_{x'}x'(\tau))<A^2_{\textrm{max}}.
\label{maximal acceleration condition}
\end{align}

There are also geometric spacetime frameworks containing a maximal proper acceleration that have a classical motivation  \cite{Schuller 2002,Gallego-Torrome2015}. Essentially, such theories violate Einstein's clock  hypothesis \cite{Einstein1922}. For such theories the spacetime metric structure is also of the form \eqref{maximalaccelerationmetric0} and the relation \eqref{maximal acceleration condition} holds good. Furthermore, in the case of metrics with a maximal acceleration in the framework described in \cite{Gallego-Torrome2015}, the relation \eqref{maximal temperature and acceleration} also holds good. Caianiello's theory, when interpreted from the point of view of high order jet geometry, is the paradigmatic example. Other theories of spacetimes with metrics of maximal acceleration that can be reformulated within the framework of high order jet geometry are the Brandt's theory \cite{Brandt1989} and Schuller's geometric framework \cite{Schuller 2002} of Born-Infeld theory \cite{Born Infeld}.

Therefore, we are interested in considering geometric frameworks whose metric structures are of the form \eqref{maximalaccelerationmetric0}, independently if the maximal acceleration has a classical or a quantum motivation. For such theories, we argue that the expression for the Unruh temperature  \eqref{Unruh temperature} is still valid. Lacking of a quantum field theory in curved spacetimes on spacetimes with metrics of maximal acceleration of the form \eqref{maximalaccelerationmetric0}, we are forced to rely on heuristic arguments to justify the validity of the Unruh temperature formula. Indeed, as discussed in \cite{Benedetto Feoli}, for the case of uniform accelerated observers, the derivation of the Unruh temperature by Alsing and Milonni \cite{Alsing Milonni} can be extended to the case when the spacetime structure is of the form \eqref{maximalaccelerationmetric0}. Such an extension is rather natural, if we consider that the coordinate transformation from an accelerated observer to an instantaneous Lorentz coordinate system is a conventional time dependent Lorentz transformation and that the dispersion relations for metrics of maximal acceleration \eqref{maximal acceleration condition} are the relativistic dispersion relations, as it was shown in \cite{Gallego-Torrome2015}. Since the derivation of Alsing and Milonni is intrinsically based upon these two ingredients in the form of the treatment of time-dependent frequency Doppler-shift, the argument also applies to spacetimes of maximal acceleration with a metric of the form \eqref{maximal acceleration condition}.

Therefore, when the spacetime structure is of type \eqref{maximalaccelerationmetric0}, there are reasons to consider that Unruh temperature formula is still valid. In this case, there is also maximal temperature given by the expression \eqref{maximal temperature and acceleration}.
\section{Weak equivalence principle in spacetimes of maximal acceleration}
One of the fundamental properties of the spacetimes $(M,g)$, where the metric structure $g$ is of the type of maximal acceleration \eqref{maximalaccelerationmetric0}, is its compatibility with a formulation of the weak equivalence principle. The weak principle of equivalence in the contest of high order jet geometries can be formulated in analogous terms as for relativistic spacetimes and as for generalized Finsler spacetimes \cite{Ricardo2017},
\\
\\
{\bf Weak equivalence principle}. {\it In the presence of a gravitational field only, under the same initial conditions in position and velocity, every test particle has the same spacetime world line as parameterized by the proper time of the limit metric $\eta$.}
\\
\\
In order to implement this principle for spacetimes with metric structure given by \eqref{maximalaccelerationmetric0}, let us consider a {\it Berwald type connection} on the second jet bundle $\pi_2:J^2_0(M)\to \,M$, where in natural local coordinates, a point in $J^2_0(M)$ has associated the coordinates $(\gamma^\mu(0),(\gamma')^\mu(0),(\gamma'')^\mu(0)),\,\mu=0,1,2,3)$, for a given curve $\gamma:I\to M$ with $0\in\,I\subset \,\mathbb{R}$. By assumption, such a connection $\nabla$ is determined by the covariant derivatives $\nabla_{{\gamma}'}\,Z$, where $\gamma '$ is a tangent vector at an arbitrary point $\gamma(0)\in M$ and $Z\in\,T\Gamma M$ viewed as the analogous elements of the associated jet bundle $J^2_0(M)$; all other covariant derivatives are set to be identically zero. These conditions on $\nabla$ when acting on holonomic basis can be stated as a {\it symmetric condition} for the connection $\nabla$, similarly as it happens in Finsler geometry for several Finslerian connections \cite{BaoChernShen}. In addition, in a local holonomic frame of $J^2_0(M)$, the non-zero connection coefficients of $\nabla$ defined by the expression $\nabla_{\frac{\partial}{\partial x^\nu}}\,\frac{\partial}{\partial x^\rho}=\,\gamma^\mu_{\nu\rho}\,\frac{\partial}{\partial x^\mu}$ are defined by the {\it Christoffel symbols}
\begin{align}
\gamma^\mu_{\nu\rho}(x,{x}',{x}'')=\,\frac{1}{2}\,g^{\mu\sigma}\,\left(\frac{\partial g_{\sigma\rho}}{\partial x^\nu}+\,\frac{\partial g_{\nu\sigma}}{\partial x^\rho}-\,\frac{\partial g_{\nu\rho}}{\partial x^\sigma}\right).
\label{connection coefficients}
\end{align}

In Fermi coordinates of $\eta$, one has $g_{\mu\nu}=\,(1-\,\frac{{x}''^\mu\,{x}''_\mu}{A^2_{\textrm{max}}})\,\eta_{\mu\nu}$, leading to the relations $\gamma^\mu_{\nu\rho}(x,{x}',{x}'')=0$. Therefore, the connections coefficients $\gamma^\mu_{\nu\rho}(x)$ coincide with the connection coefficients of the Levi-Civita connection of $\eta$ in any local holonomic frame, by the transformation law of the connection coefficients.
 In this setting, the weak equivalence principle is implemented by stating that the world line of free falling test particles are auto-parallel curves of the connection $\nabla$ as parameterized curves where the parameter is the proper time of $\eta$ along $\gamma:I\to M$. Therefore, the world lines of free falling test particles satisfy the condition
\begin{align}
\nabla_{\gamma '(t)}\,\gamma ' (t) =0 ,
\label{autoparallel condition of nabla}
\end{align}
This auto-parallel condition coincide with the auto-parallel curves of $D$ as affine parameterized curves,
\begin{align}
D_{\gamma '(t)}\,\gamma ' (t) =0,
\label{geodesics of D}
\end{align}
since the connection coefficients of $\nabla$ coincide with the connection coefficients of $D$, as it is easily checked. It is direct that the weak equivalence principle as stated above holds good for this characterization of the free-fall. Therefore, we are in a situation where, locally, a gravitational field is indistinguishable of inertial acceleration.

\section{Maximal acceleration implies a minimal mass for Schwarzschild's type black holes}
 The Bekestein-Hawking temperature of the relativistic Schwarzschild black hole is given by the expression \cite{Hawking 1974}
\begin{align}
T_{BH}=\frac{\hbar\,\kappa}{2\,\pi\,c\,k_B}=\,\frac{\hbar\,c^3}{8\,\pi\,G\,k_B\, M},
\label{Hawking temperature}
\end{align}
where $M$ is the black hole mass and $\kappa$ is the surface gravity, that for a Schwarzschild black hole is given by $\kappa =\,\frac{c^4}{2M\,G}$. We seek to generalize this relation to the framework of high order jet spacetimes of  maximal acceleration. In order to achieve such a generalization we rely on the fact that Hawking's radiation and with it, Hawking temperature, can be seen as a consistent condition between the Unruh effect near the black hole horizon and the weak equivalence principle. Since Unruh temperature formula and the weak equivalence principle holds for the spacetimes that we are interested in, this support the view that the expression for the Bekenstein-Hawking temperate must also be valid for the theories with metrics of maximal acceleration that we are considering, as long as the surface curvature of the Schwarzschild's type black hole in spacetimes of maximal acceleration remains the same at first order approximation. That this is the case can be seen from the corrections to Schwarzschild black hole metric coming from a maximal acceleration geometry, as discussed in \cite{FeoliLambiasePapiniScarpetta1997}, where the corrections to the spacetime metric arise as a function of $\frac{1}{A_{\textrm{max}}}$. The same dependence is transported to the surface gravity, given by analytic expressions of the metric\footnote{Let us here remark that this type of corrections due to the effects of maximal acceleration are present for other relevant solutions of Einstein equations, as discussed in the context of Caianiello's theory in Reissner-Norstrom and Kerr solutions \cite{BozzaFeoliPapiniScarpetta, BozzaFeoliLambiasePapiniScarpetta}. Furthermore, such corrections are consistent with experimental astrophysical constrains in several situations \cite{Bozza et al. 2001, Papini et al. 2002, Papini 2002}, quantum dynamics in superconductors \cite{Papini 2002 b}atomic physics\cite{Lambiase et al. 1998} and classical electrodynamics \cite{Feoli et al. 1997}.}. From the definition of surface gravity as a limit acceleration for observers far away from the black hole region, the relation $|A_{\textrm{max}}|\geq \,|\kappa|$ must hold. In particular, for a Schwarzschild black hole, Hawking's temperature \eqref{Hawking temperature} follows from Unruh temperature (redshifted from the black hole region to an asymptotically far region) and is valid as a first approximation in $\kappa/A_{\textrm{max}}$. Then the bound $T_{BH}<\,T_{\textrm{max}}$ applies. This implies a minimum mass for the black hole given by
\begin{align}
M_{\textrm{min}}=\,\frac{c^4}{4\,G}\, A^{-1}_{\textrm{max}}.
\label{minimal mass for Schwarzschild}
\end{align}
Besides the possible quantum origin of the maximal proper acceleration $A_{\textrm{max}}$, the minimum mass $M_{\textrm{min}}$ of the Schwarzschild black hole depends on the classical physics constants $c$ and $G$ and is independent from the Planck's constant or other quantum related physical constants. Also, the minimal mass is independent of the initial mass of the black hole.

It is remarkable that relation \eqref{minimal mass for Schwarzschild} establishes a duality between two very different objects, the black hole (represented by the mass $M_{\textrm{min}}$ and the test particle (that can experience the acceleration limited by $A_{\textrm{max}}$).

Expression \eqref{minimal mass for Schwarzschild} implies that, when there is a maximal acceleration of the types discussed above, a black hole of Schwarzschild type will not fully evaporate by emission of Hawking radiation in spacetimes. This conclusion holds for theories considered above. It also holds for theories where the justification of the maximal acceleration is quantum mechanical or based on quantum gravity arguments, but the spacetime arena is Lorentzian and the weak equivalence principle holds.

The scale of the maximal acceleration is essential. Since it depends greatly on the theory considered, in models where the maximal acceleration is large but not as large as the Planck acceleration, the mass of the remnant appears to be relatively large. This makes plausible that the structure can  encode the mixing information of the initial degrees of freedom.
\section{Examples}
We can illustrate the above discussion with several examples.
For Brandt's theory and theories of maximal proper acceleration based upon models of quantum gravity, the scale of the maximal acceleration is the Planck acceleration $A_{\textrm{max}}\sim\,\left(c^7/\hbar\,G\right)^{1/2}$, which is of order $10^{52}\,ms^{-2}$. This value of the maximal acceleration is equivalent to a $M_{\textrm{min}}$ of order of the Planck mass $M_{Pl}=\,\left(\hbar \,c/G\right)^{1/2}\sim\,10^{-38}\,\textrm{M}_{\odot}$. This result has been already discussed in the literature \cite{Brandt1983}.

 In the case of
 Caianiello's theory, the maximal proper acceleration is given by the expression
 \begin{align*}
 A_{\textrm{Cai}}=\,\frac{2\,m\,c^3}{\hbar}
  \end{align*}
  For the electron, Caianiello's theory provides a maximal acceleration of order $A_{\textrm{Cai}}\sim \,\,10^{29}\,m/s^2$ and the associated black hole minimal mass of order $M_{\textrm{min}}\sim \,10^{-15}\, \textrm{M}_{\odot}$, which is a much smaller scale than for Brandt's theory. For other elementary  particles, the associated maximal acceleration is larger, since it depends on the mass $m$ linearly. For instance, for a proton we will have $M_{\textrm{min}}\sim \,10^{-18}\, \textrm{M}_{\odot}$, where $\textrm{M}_{\odot}$ is the solar mass.

Let us now consider theories for which the maximal proper acceleration is associated with a classical theory. For the Born-Infeld non-linear electrodynamics the metric structure of spacetime is a Lorentzian metric. The theory contains a maximal proper acceleration given by
\begin{align*}
A_{BI}=\,\frac{q}{m}\,b^{-1},
\end{align*}
 where $q$ and $m$ are the charge and mass of a test particle approaching the horizon of the black hole and $b^{-1}$ is an universal constant, the Born-Infeld constant. The parameter $b$ is not fixed by the theory and currently there are only upper bounds on it \cite{Schuller 2002}. Since the spacetime is Lorentzian, one can apply Unruh's formula directly. Then the minimum mass of a Schwarzschild black hole in Born-Infeld theory is
\begin{align}
M_{\textrm{min}}=\,\frac{c^4}{4\,G}\,\frac{m_e}{e}\,b,
\label{Born Infeld maximal acceleration}
\end{align}
where $e$ and $m_e$ are the charge and mass of the electron. Experimental bounds on $b$ provide lower bounds on $M_{\textrm{min}}$. Current bounds on $b$ imply the bound $M_{\textrm{min}}\leq \, 10^{-5}\,\textrm{M}_{\odot}$.

Another example of classical theory with a maximal proper acceleration is Caldirola's theory \cite{Caldirola1981}, whose maximal acceleration is of the same order as higher order classical electrodynamics \cite{Gallego-Torrome2019}. For both theories, the maximal proper acceleration is of order $A_{\textrm{max}}=\,\frac{3}{2}\,\epsilon_0\,\frac{m\,c^4}{q^2}$, where $\epsilon_0$ is the dielectric constant of the vacuum. For an electron, this maximal acceleration is of order $10^{32}\,ms^{-2}$, which is equivalent to a minimal mass for black holes of Schwarzschild type of the form $M_{\textrm{min}}\sim \, 10^{-18}\,\textrm{M}_{\odot}$. On the other hand, the relevant acceleration in quantum electrodynamics is the critical acceleration \cite{Schwinger1951,di Piazza et al.}, of order $a_{cr}\sim \,10^{-29}\,ms^{-2}$ which is equivalent to a black hole minimal mass  $M_{\textrm{min}}\sim\,10^{-15}\textrm{M}_{\odot}$.

The above example show that the remnant can be relatively large in mass.
\section{Relation with the conjecture of maximum force}
The relation \eqref{minimal mass for Schwarzschild} has some consequences for the conjecture of maximum force in general relativity \cite{Gibbons 2002,Barrow Gibbons, Schiller 2005}. We state the conjecture as saying that in general relativity, the maximum force exerted between two bodies is bounded by $c^4/4G$. Since in general $m\,a\leq \,m\,A_{\textrm{max}}$, if one assumes that the mass $m$ of the test particle is always less than the minimal mass  $M_{\textrm{min}}$ attainable by a black hole, then
\begin{align}
m\,a\leq\,\frac{c^4}{4G}.
\label{maximal force}
\end{align}
Thus in presence of a maximal acceleration and via the relation \eqref{minimal mass for Schwarzschild}, there is a maximal force that can be exerted on any test particle with $m<\,M_{\textrm{min}}$.

A closer analysis shows that indeed, the relation \eqref{maximal force} as devised in the above framework is different from the conjecture of maximum acceleration, since it holds in situations when there is a finite maximal proper acceleration of the type discussed above, that excludes general relativity. Furthermore, in the limit $A_{\textrm{max}}\to +\infty$, the consistency of the relation \eqref{minimal mass for Schwarzschild} implies that the minimal of the black hole can be arbitrarily small, $M_{\textrm{min}}\to 0$, in which case there are test particles with mass $m'\gg M_{\textrm{min}}$ such that \eqref{maximal force} is violated. This conclusion holds in the case of general relativity, since in that theory there is no apparent limit to the acceleration of test particles.
\section{Conclusion}
It has been shown that in frameworks of classical spacetimes of maximal acceleration and where the weak equivalence principle holds there must exist a lower bound for the mass of the black hole of Schwarzschild type, a bound that depends on the inverse of the maximal proper acceleration. Since in principle the mass of such a remnant can be {\it large}, this result opens a new avenue to treat the information paradox problem. Note that our arguments are based on classical high order jet geometries, in contrast with the many tentative ideas of solving the information paradox, that lay on arguments where gravity is quantized.

We have also shown that in such frameworks of classical maximal acceleration geometries there is an upper bound for the force exerted on test particles. Such a limit resembles the conjectured maximum force in general relativity, but it is in principle a different result. Indeed, we argue that in the case of general relativity, when $A_{\textrm{max}}\to +\infty$, there is no uniform bound on the force experienced by test particles.

\footnotesize{
}

\end{document}